\documentclass[preprint,floatfix]{revtex4}
\usepackage{graphicx}
\usepackage{epsfig}
\usepackage{enumerate}
\usepackage{amssymb}
\usepackage{amsmath}
\usepackage{amsfonts}
\usepackage{dcolumn}
\usepackage{bm}
\usepackage{color}
\usepackage{siunitx}
\usepackage{layout}

\begin{document}
\begin{center}

\textbf{\large{Towards compact phase-matched and waveguided nonlinear optics in atomically layered semiconductors}}
\vspace{0.5cm}

Xinyi Xu$^1$, Chiara Trovatello$^{1,2}$, Fabian Mooshammer$^{3}$, Yinming Shao$^{3}$, Shuai Zhang$^{3}$, Kaiyuan Yao$^{1}$,  Dmitri N. Basov$^{3,*}$, Giulio Cerullo$^{2,*}$ and P. James Schuck$^{1,*}$

\vspace{0.5cm}

\textit{\footnotesize{$^1$ Department of Mechanical Engineering, Columbia University, New York, NY, USA\\
$^2$ Dipartimento di Fisica, Politecnico di Milano, Piazza L. da Vinci 32, I-20133 Milano, Italy \\
$^3$ Department of Physics, Columbia University, New York, NY 10027, USA\\
}}

\vspace{0.5cm}

{\footnotesize \textit{Email address:} db3056@columbia.edu, giulio.cerullo@polimi.it,  p.j.schuck@columbia.edu}

\section*{Abstract}
\end{center}
\noindent
{\footnotesize Nonlinear frequency conversion provides essential tools for light generation, photon entanglement, and manipulation. Transition metal dichalcogenides (TMDs) possess huge nonlinear susceptibilities and 3R-stacked TMD crystals further combine broken inversion symmetry and aligned layering, representing ideal candidates to boost the nonlinear optical gain with minimal footprint. Here, we report on the efficient frequency conversion of 3R-MoS$_2$, revealing the evolution of its exceptional second-order nonlinear processes along the ordinary (in-plane) and extraordinary (out-of-plane) directions. Along the ordinary axis, by measuring difference frequency and second harmonic generation (SHG) of 3R-MoS$_2$ with various thickness - from monolayer ($\sim\SI{0.65}{nm}$) to bulk ($\sim\SI{1}{\micro m}$) - we present the first measurement of the SHG coherence length ($\sim\SI{530}{nm}$) at $\SI{1520}{nm}$ and achieve  record nonlinear optical enhancement from a van der Waals material, $>10^4$ stronger than a monolayer. It is found that 3R-MoS$_2$ slabs exhibit similar conversion efficiencies of lithium niobate, but within propagation lengths that are more than $100$-fold shorter at telecom wavelengths. Furthermore, along the extraordinary axis, we achieve broadly tunable SHG from 3R-MoS$_2$ in a waveguide geometry, revealing the coherence length in such structure for the first time. We characterize the full refractive index spectrum and quantify both birefringence components in anisotropic 3R-MoS$_2$ crystals with near-field nano-imaging. Empowered with these data we assess the intrinsic limits of the conversion efficiency and nonlinear optical processes in 3R-MoS$_2$ attainable in waveguide geometries. Our analysis highlights the potential of 3R-stacked TMDs for integrated photonics, providing critical parameters for designing highly efficient on-chip nonlinear optical devices including periodically poled structures, resonators, compact optical parametric oscillators and amplifiers, and optical quantum circuits.\\}

\noindent Nonlinear optics lies at the heart of light generation and manipulation. Coherent frequency conversion, such as second- and third-harmonic generation, parametric light amplification and down-conversion, enables a deterministic change in wavelength as well as control of temporal and polarization properties. When integrated within photonic chips, nonlinear optical materials constitute the basic building blocks for all-optical switching\cite{1,2,2-1}\noindent, light modulators\cite{3,3-1,3-2,3-3}\noindent, photon entanglement\cite{4,5} and optical quantum information processing\cite{6, 6-1}\noindent. Conventional nonlinear optical crystals display moderate second-order nonlinear susceptibilities ($|\chi^{(2)}|\sim1-\SI{30}{pm/V}$) and perform well in benchtop setups with discrete optical components. However, such crystals do not easily lend themselves to miniaturization and on-chip integration. Two-dimensional transition metal dichalcogenides (TMDs) possess huge nonlinear susceptibilities\cite{7} ($|\chi^{(2)}|\sim100-\SI{1000}{pm/V}$) and, thanks to their deeply sub-wavelength thickness, offer a unique platform for on-chip nonlinear frequency conversion\cite{8} and light amplification\cite{9}\noindent. Furthermore, their semiconducting properties render TMDs superior for applications compared to opaque materials with exceptionally large $|\chi^{(2)}|$ such as Weyl semimetals\cite{10}\noindent.

In single- or few-layer TMD samples, SHG is extensively exploited for characterization of structural properties such as crystal orientation\cite{11,11-1,11-2,11-3} or local strain\cite{12}\noindent. However, due to their atomic thickness, these samples display a notably lower SHG efficiency ($\eta_{SHG} = I_{2\omega}/I_{\omega}\sim 10^{-11}$ at $I_{\omega} = \SI{30}{GW/cm^2}$) compared to standard nonlinear crystals ($\eta_{SHG} = I_{2\omega}/I_{\omega}\sim 1-50\%$). The SHG efficiency can be written as\cite{13}\noindent: $\eta_{SHG} \propto |\chi^{(2)}|^2L^2$, where $L$ is the thickness of the nonlinear medium (assuming \textbf{}perfect phase matching and non-depletion regime). The nonlinear conversion efficiency of a TMD could thus be scaled by increasing the propagation length $L$ through the active medium. This is attainable by increasing the number of layers in the TMD sample. However, the nonlinear optical properties of multilayer TMDs critically depend on their crystallographic symmetry\cite{14}\noindent. 

Group VI trigonal TMDs (e.g. MoS$_2$) are stable in two crystallographic phases: polytype 2H (hexagonal) and polytype 3R (rhombohedral)\cite{15}\noindent. 2H-MoS$_2$ is naturally centrosymmetric, giving an opposite dipole orientation among consecutive layers. This results in a vanishing nonlinear susceptibility ($|\chi^{(2)}| = 0$) for crystals with even number of layers\cite{16,11} and precludes efficient conversion in multilayer 2H-TMDs. To circumvent this limitation – and restore the quadratic scaling of the nonlinear conversion efficiency with the number of layers N ($I_{2\omega}/I_{\omega} \propto N^2$) – one can artificially AA stack several monolayers\cite{9,14}\noindent, aligning their dipole moments\cite{13,13-1}\noindent. Although the mechanically assembled stacks serve as proof of concept for fundamental studies, their labor-intensive fabrication prevents massive large-scale production. 

In contrast, 3R-MoS$_2$ is naturally non-centrosymmetric. The optical emission from consecutive in-plane nonlinear dipoles of 3R-MoS$_2$ results in a constructive interference, prompting the $N^2$ enhancement of the nonlinear conversion efficiency\cite{9,14} for thin samples. Similar to 2H-MoS$_2$, bulk 3R-MoS$_2$ can be grown by chemical vapor transport (CVT)\cite{17}\noindent and thin 3R-MoS$_2$ flakes can be obtained by dry mechanical exfoliation. The nonlinear optical response of 3R-MoS$_2$ has been explored in some recent pioneering studies, so far focusing on thinner crystals, reporting the $N^2$ enhancement at the 2D limit, and showing a maximum SHG enhancement of $\sim10^2$ occurring at specific thickness windows\cite{17,18}\noindent. Pushing towards general application, however, requires higher nonlinear enhancements and thus larger $N$, which in turn leads to more intricate interferences and interactions within the crystal. Specifically, for multilayer TMDs, the wavevector mismatch between the fundamental wavelength (FW) and the second harmonic (SH) needs to be considered, as it limits the maximum propagation length for constructive interference. In addition, thick 3R-MoS$_2$ crystals act as Fabry-Perot cavities, which modulate the FW power inside the sample. The combination of these effects determines the optimum thickness of 3R-MoS$_2$ for the highest SHG conversion efficiency. Due to their layered nature, 3R-stacked TMDs are also naturally anisotropic, and thus birefringent – a key prerequisite for achieving perfect phase-matching.

Here we measure SHG and difference frequency generation (DFG) from multilayer 3R-MoS$_2$ crystals with variable thickness, using a custom transmittance microscope to determine the maximum enhancement of nonlinear conversion efficiency, revealing the intrinsic upper limits of the material. We provide a comprehensive model, which explains the second-order nonlinearity of 3R-MoS$_2$ including its phase mismatch and its intrinsic interference effects. We report the first measurement of the coherence length $L_c$ of 3R-MoS$_2$, elucidating the role of phase-matching at excitation photon energies close to the telecom band. In addition, we reveal that 3R-MoS$_2$ enables broadband SH conversion in waveguide geometries. Upon edge coupling of the FW, we detect and map both FW and SH emission from the opposite edge of the flake within our field of view. We observe the characteristic SHG signal modulation with increasing path length, allowing us to quantify the out-of-plane coherence length in 3R waveguide structures. Further, we also characterize the anisotropic linear optical properties by imaging the propagation of waveguide modes in real space using near-field nano-imaging, identifying the conditions for phase-matched SHG in waveguide geometries. Together, these findings pave the way for achieving birefringent phase matching in waveguides of van der Waals (vdW) semiconductors, directly impacting the field of vdW photonics by enabling future advances in conversion efficiencies and integration.\\

\section*{Results}

\begin{figure}[h!]
\centering
\includegraphics[width=1\textwidth]{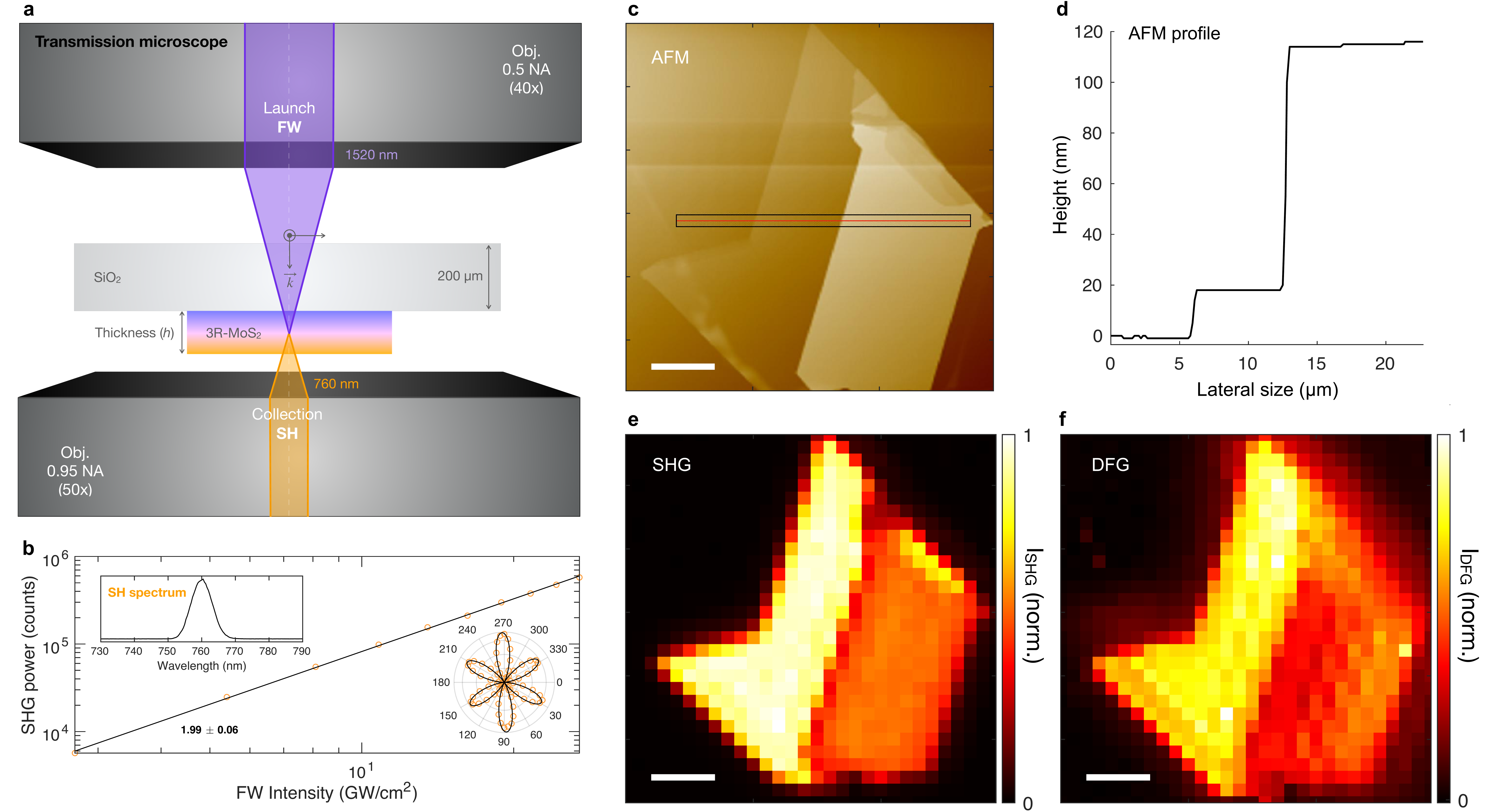}
\caption{\scriptsize{\textbf{SHG and DFG emission from 3R-MoS$_2$.} \textbf{a,} Schematic diagram of the transmittance microscope. Excitation of 3R-MoS$_2$ with thickness $h$ is through a 40x reflective objective (NA=0.5) and the nonlinear emission is collected by a 50x objective (NA=0.95). The sample is exfoliated on a transparent $\SI{200}{\micro m}$ thick SiO$_2$ substrate. \textbf{b,} SH intensity as a function of FW power. Insets: (top left) representative SH spectrum, (top down) polar plot of the armchair directions. \textbf{c,} AFM image of a representative 3R-MoS$_2$ flake. \textbf{d,} AFM profile of the marked region. The flake includes two uniform regions of 20 and $\SI{119}{nm}$. \textbf{e,} Normalized SHG map at $\SI{1.63}{eV}$. The FW photon energy is $\SI{0.815}{eV}$. At each data point the pump power is kept constant at $\SI{5.4}{mW}$ and the linear pump polarization is parallel to the armchair direction. \textbf{f,} Normalized DFG map at $\SI{2.16}{eV}$. The pump photon energy is $\SI{3.11}{eV}$ and the signal photon energy is $\SI{0.95}{eV}$. At each data point the pump and the signal powers are $\SI{121}{\micro W}$ and $\SI{93}{mW}$, respectively. Pump and signal have the same linear polarization, parallel to the armchair direction. The scale bar is $\SI{5}{\micro m}$.}}
\label{fig:1}
\end{figure}

\noindent We use a custom transmission microscope (see Methods) (Fig. \ref{fig:1} a) to measure SHG and DFG from the multilayer 3R-MoS$_2$ flakes with tunable thickness $h$. The 3R-MoS$_2$ micro-crystals are mechanically exfoliated from a commercial CVT-grown bulk 3R-MoS$_2$ crystal (HQ graphene) onto a $\SI{200}{\micro m}$ thick fused silica (SiO$_2$) substrate. The bulk sample has been characterized by energy dispersive X-ray analysis (EDX) and X-Ray diffraction (XRD) (Supplementary Figure 1). The thickness of each exfoliated flake has been determined by atomic force microscopy (AFM), see Supplementary Note 1, and Supplementary Figures 2 and 3. The detection objective has a larger numerical aperture (NA) than the excitation one to maximize signal collection from scattering at larger angles. 

Figure \ref{fig:1}b shows power-dependent SHG measured on $\SI{119}{nm}$ thick 3R-MoS$_2$ (dots) and the fitted power law (line). The pump wavelength is set to $\SI{1520}{nm}$ ($\SI{0.815}{eV}$) yielding SHG centered at $\SI{760}{nm}$ ($\SI{1.63}{eV}$) (see inset for a representative spectrum). The SHG emission follows the expected quadratic power dependence. The saturation regime is beyond the maximum excitation power that we can achieve at the focus in our setup, i.e., $\sim\SI{45}{mW}$, corresponding to an intensity of $\sim\SI{120}{GW/cm^2}$. Moreover, since both FW and SH are tuned below the bandgap of 3R-MoS$_2$, the material is essentially transparent, and no appreciable degradation of the sample is detected (see Supplementary Figure 6). This highlights the potential to boost the nonlinear conversion efficiency at higher intensities. Due to damage considerations, such intensities are usually unattainable in the absorptive above-gap regime, where excitonic resonances are exploited to enhance the nonlinear response of TMDs\cite{7,11-2}\noindent. A representative 6-lobed polarization-dependent SHG flower pattern\cite{11-1}(Fig. \ref{fig:1} b inset), in which the pump polarization is rotated by a half-wave plate, and the transmission axis of the detection polarizer is kept parallel to the pump, reflects the $D_\mathrm{6h}$ point group of the 3R crystal with broken inversion symmetry. It shows two longer lobes along one of the armchair directions, attributable to the staggered stacking direction\cite{17}\noindent. 

\begin{figure}[h!]
\includegraphics[width=1\textwidth]{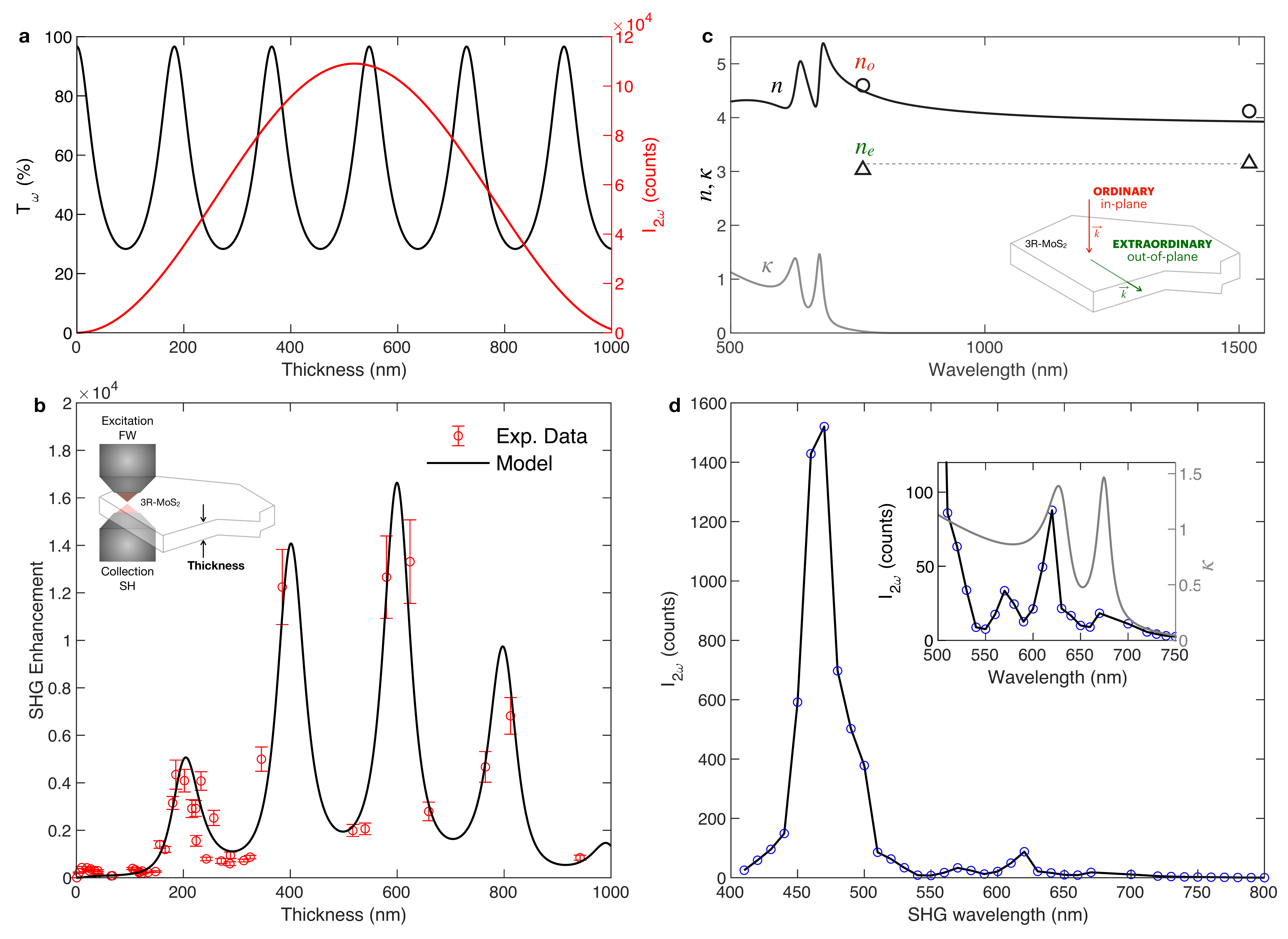}
\centering
\caption{\scriptsize{\textbf{In-plane SHG coherence length.} \textbf{a,} Calculated pump transmissivity (black) and phase-mismatch curve (red) as a function of the 3R-MoS$_2$ thickness. \textbf{b,} Measured thickness-dependent SHG enhancement of 3R-MoS$_2$ with respect to the monolayer (circles) and calculated theoretical enhancement (line). FW and SH photon energies are $\SI{0.815}{eV}$ and $\SI{1.63}{eV}$, respectively. The pump power is kept constant at $\SI{5.4}{mW}$ and the linear pump polarization is parallel to the armchair direction. The error bars represent the variance of the nonlinear signal over the flake area, originating from the sample inhomogeneity, which induces a fluctuation in the nonlinear signal of $\sim10\%$. \textbf{c,} Real ($n$) and imaginary ($\kappa$) part of the refractive index of bulk 3R-MoS$_2$ ($n_1 = n + i\kappa$). $n$ and $\kappa$ are extracted from the transmittance and reflectance spectra (see Supplementary Note 3) of a representative $\SI{94}{nm}$ thick 3R-MoS$_2$ on fused silica substrate. The peaks in $\kappa$ at $\SI{675}{nm}$ and $\SI{624}{nm}$ are attributed to A and B excitonic resonances. Circles and triangles represent the ordinary ($n_o$) and extraordinary ($n_e$) refractive indexes determined by s-SNOM, illustrated in Fig. \ref{fig:5}. The dashed line indicates the average $n_e$ in the low energy range, expected to be nearly constant\cite{20}\noindent. \textbf{d,} SHG excitation spectrum measured on a $\SI{4.2}{nm}$ thick 3R-MoS$_2$, with a constant pump power of $\SI{1.35}{mW}$ and tunable FW ($\SI{1.55}{eV}-\SI{3.02}{eV}$). Inset: Comparison between the SHG spectrum and the imaginary refractive index $\kappa$ (grey line) zooming in the excitonic resonance absorption energy range.}}
\label{fig:2}
\end{figure}

Figure \ref{fig:1}c shows the AFM image of a representative 3R-MoS$_2$ flake, along with a line cut of the height profile (Fig. \ref{fig:1}d), in which we can distinguish two flat regions of $\SI{20}{nm}$ and $\SI{119}{nm}$ thickness. A sample-scanning confocal modality is used for mapping the spatially dependent SHG and DFG intensities over the flake (Fig. \ref{fig:1}e and \ref{fig:1}f, respectively). The SHG (FW at $\SI{1520}{nm}$, $\SI{0.815}{eV}$) is measured with the pump polarization and the collection analyzer directions parallel to the armchair direction with the largest nonlinear response. In Fig. \ref{fig:1}e, the $\SI{20}{nm}$ thick region displays an SHG intensity twice as large as the one obtained on a $\SI{119}{nm}$ thick flake. In other words, by increasing the thickness of the 3R-MoS$_2$ flake, the emitted SHG decreases. Since both FW and SH photon energies lie below the direct optical bandgap ($\sim\SI{1.85}{eV}$), this effect cannot be attributed to absorption (indirect absorption losses are negligible for these wavelengths and thicknesses, see Fig. \ref{fig:2}c).

The DFG map at $\SI{574}{nm}$ ($\sim\SI{2.16}{eV}$), shown in Fig. \ref{fig:1}f, is recorded on the same flake using a pump wavelength of $\SI{400}{nm}$ ($\sim\SI{3.11}{eV}$) and a signal at $\SI{1300}{nm}$ ($\sim\SI{0.95}{eV}$). The pump and signal beams have parallel polarizations, while the collection is unpolarized. Note that, as with the SHG, the thicker area has weaker DFG signal than the thinner area. Considering that both pump and idler photon energies lie above the optical gap, we estimate that the absorption is the main reason for measured weaker idler intensity in this case. Indeed, we cannot detect any idler signal through a $\SI{622}{nm}$ thick 3R-MoS$_2$ flake.

To understand the thickness-dependence of the SHG efficiency, we must take into account both interference and phase-matching effects. We analyze the light propagation in the nonlinear medium using the transfer matrix method (TMM), modeling our structure as a 3-layer system SiO$_2$/MoS$_2$/air with refractive indexes n$_0$/n$_1$/n$_2$. The transmissivity of the FW light changes periodically with the sample thickness $h$ (see Supplementary Note 2 for the extended calculation) as: 
\begin{equation}
    T_{\omega} (h)= \frac{Re\{n_{2}\}}{Re\{n_{0}\}}\left |\frac{t_{01} t_{12}}{e^{jk_{1}h} + r_{01}r_{12}e^{-jk_{1}h}}  \right |^2\label{(1)}
\end{equation}
\noindent
Where $n_i$ is the refractive index of each layer, $t_{ij}$ and $r_{ij}$ are the transmissivity and reflectivity coefficients from layer $i$ to layer $j$, $k$ is the wavevector, and $h$ is the thickness of the 3R-MoS$_2$ layer. The effective FW intensity at the sample is $I_{\omega,s}= T_{\omega}(h)I_{\omega,in}$ where $I_{\omega,in}$ is the FW intensity after the focusing objective, which is kept fixed during the experiment. Due to interference effects, the effective power flux across the sample will change periodically along with the thickness (black curve, Fig. \ref{fig:2}a). 

The discrepancy in refractive index for the FW at frequency $\omega$ and the SH at $2\omega$ sets further constraints on conversion. Efficient frequency conversion in bulk nonlinear crystals is achieved by fulfilling the phase-matching condition, i.e. by coherently adding the signals generated at different longitudinal coordinates of the crystal. Due to the frequency dependence of the refractive index, after a certain propagation length the locally generated SH will be out of phase with the SH from previous planes of the crystal. The overall SH intensity continues to grow until the so-called coherence length $L_c$ is reached and then begins to decrease due to destructive interference\cite{13}\noindent. The SH intensity under phase-mismatched conditions can be written as:
\begin{equation}
    I_{2\omega}\propto\frac{|\chi^{(2)}|^2}{\Delta k^2}I_{\omega}^{2}sin^2(\frac{\Delta kh}{2})
\end{equation}
\noindent
where $\Delta k = k_{2\omega} - 2k_{\omega} = 2\omega/c(n_{2\omega}-n_{\omega})$ is the wavevector mismatch between the SH and the FW (red curve, Fig. \ref{fig:2}a). Equation (2) shows that the maximum efficiency is reached for a thickness of the nonlinear crystal corresponding to the coherence length $L_c = \pi/\Delta k$. Combining thickness-dependent FW transmission $T_{\omega}(h)$ and the phase-matching relationship, one can see that the SHG efficiency is modulated by both multilayer interference effects and by the phase mismatch, giving an optimal thickness of the nonlinear crystal.

As noted above, to avoid absorption losses, we choose FW and SH photon energies below the optical gap of MoS$_2$. The experimental data of the measured nonlinear emission and the fitting curve $I_{2_{\omega}} (h)$ are shown in Fig. \ref{fig:2}b, with the amplitude as the only free fitting parameter. The  $\sim10\%$ fluctuation of the nonlinear signal originates from the sample spatial inhomogeneity. The measured real refractive indices of 3R-MoS$_2$ are $n_{\omega} = 3.795$ at $\SI{0.815}{eV}$ and $n_{2\omega} = 4.512$ at $\SI{1.63}{eV}$, and the corresponding real refractive index mismatch is $n_{2\omega} - n_{\omega}= 0.717$, which is in agreement with previously reported values for bulk 2H-MoS$_2$\cite{19,20}\noindent. These values give, for a pump photon energy of $\SI{0.815}{eV}$, a coherence length $L_c\sim \SI{530}{nm}$ and a transmittance period of $\SI{182}{nm}$ for 3R-MoS$_2$, in excellent agreement with experimental results (Fig. \ref{fig:2}b). In the low thickness regime, the deviation of the experimental data from the model calculated with the TMM is due to the evolution of the  bandstructure. The refractive index of mono- and few-layer TMDs differs from the refractive index of bulk MoS$_2$\cite{20-1}\noindent, with thinner films having smaller refractive index and larger overall transmissivity. In our model, we estimate the thickness-dependent SHG using the bulk refractive index. Therefore, at lower thicknesses, the SHG intensity will be higher than the calculated one.

The largest experimental SHG enhancement with respect to a monolayer, obtained for a $\SI{622}{nm}$ thick 3R-MoS$_2$ crystal, is approximately $1.5\times10^4$. Ideally, covering the flake with an anti-reflection coating at the FW could further increase the nonlinear conversion efficiency. According to the phase mismatching curve (red line in Fig. \ref{fig:2}a), choosing a 3R-MoS$_2$ thickness of $\SI{530}{nm}$ would yield the intrinsic limit for enhancement of $\sim1.1\times10^5$ times with respect to the monolayer MoS$_2$ within one coherence length at the pump photon energy of $\SI{0.815}{eV}$. Considering that the reported conversion efficiency of monolayer MoS$_2$ at FW = 1560 nm is $\sim 7\times10^{-11}$ at $\SI{30}{GW/cm^2}$\cite{16}, the overall conversion efficiency of MoS$_2$ at the coherence length thickness will be $\sim 10^{-6}-10^{-5}$. Our results show that, in order to realize an optimal nonlinear conversion efficiency, one needs to choose a material thickness close to the coherence length and that at the same time guarantees constructive interference for the FW. Further enhancement can then be achieved by regularly structuring or poling larger crystals or waveguides with a periodicity on this length scale, or by exploiting birefringence. 

The advantage of 3R-MoS$_2$ for nonlinear frequency conversion becomes particularly striking when one compares its conversion efficiency density $\eta := P_{SH}/(P_{FW}^2L^2)$ with that of state-of-the-art LiNbO$_3$ devices at the telecom wavelength.  Utilizing our measured material parameters, we calculate $\eta = \SI{71800}{\%W^{-1}cm^{-2}}$ in 3R-MoS$_2$ for $L = \SI{622}{nm}$, while $\eta = \SI{460}{\%W^{-1}cm^{-2}}$ for LiNbO$_3$ on an insulator waveguide with $\SI{50}{\micro m}$ propagation length\cite{21}\noindent. The coherence length $L_c$ of LiNbO$_3$ at FW $\SI{1545}{nm}$ is $\SI{9.5}{\micro m}$\cite{22}\noindent, and the conversion efficiency at the coherence length $L_c$ is $I_{2\omega}/I_{\omega}\sim 3\times 10^{-8}$. Notably, 3R-MoS$_2$ achieves similar conversion efficiencies with two orders of magnitude shorter propagation lengths.

To probe the effects of excitonic and interband transitions on the $\chi^{(2)}$ of  3R-MoS$_2$, we obtained the full refractive index spectrum of a bulk crystal using a combination of transmission and reflection experiments and compare the results with the SHG frequency dependence. We report the full refractive index spectrum for in-plane polarization in Fig. \ref{fig:2}c (for a wide range spectrum see SI).  The real and imaginary components of the index, $n$ and $\kappa$, are retrieved from the complex dielectric function $\epsilon$, which is extracted from transmittance (T) and reflectance (R) spectra measured on a $\sim\SI{94}{nm}$ 3R-MoS$_2$ crystal on a fused silica substrate (see SI). The absorption resonances of the $\kappa(\lambda)$ spectrum, highlighted in the inset of Fig. \ref{fig:2}d, are attributed to excitonic effects. In particular, the peaks at $\SI{675}{nm}$ and $\SI{624}{nm}$ are A and B excitons\cite{23,41}\noindent. The onset of the transparency region of 3R-MoS$_2$ lies at $\sim\SI{750}{nm}$. 

Figure \ref{fig:2}d shows the SHG spectrum measured on a $\SI{4.2}{nm}$ thick 3R-MoS$_2$ flake on $\SI{200}{\micro m}$ thick SiO$_2$, revealing the wavelength dependence of the $\chi^{(2)}$ of 3R-MoS$_2$ along the armchair direction. The response of our system has been calibrated with a standard alpha-quartz sample. The error of the measurement is negligible, as it mainly originates from the laser power fluctuations, inducing a change in the nonlinear signal of $\sim0.1\%$. Here, each point results from the average of 10 integrated spectra measured on a single spot of the flake. The main peaks at $\sim 670$ and $\SI{620}{nm}$ are consistent with the A and B exciton absorption resonances\cite{24} measured on bulk 3R-MoS$_2$ ($\kappa$ spectrum in grey), while the peak at $\SI{470}{nm}$ originates from high-energy transitions at the band nesting region between $K$ and $\Gamma$ points of the Brillouin zone. The slight energy deviation from the excitonic resonances in 2H-MoS$_2$ can be attributed to the different crystal structure of the 3R polytype affecting the band structure and the optical absorption.\\

\begin{figure}[h!]
\includegraphics[width=1\textwidth]{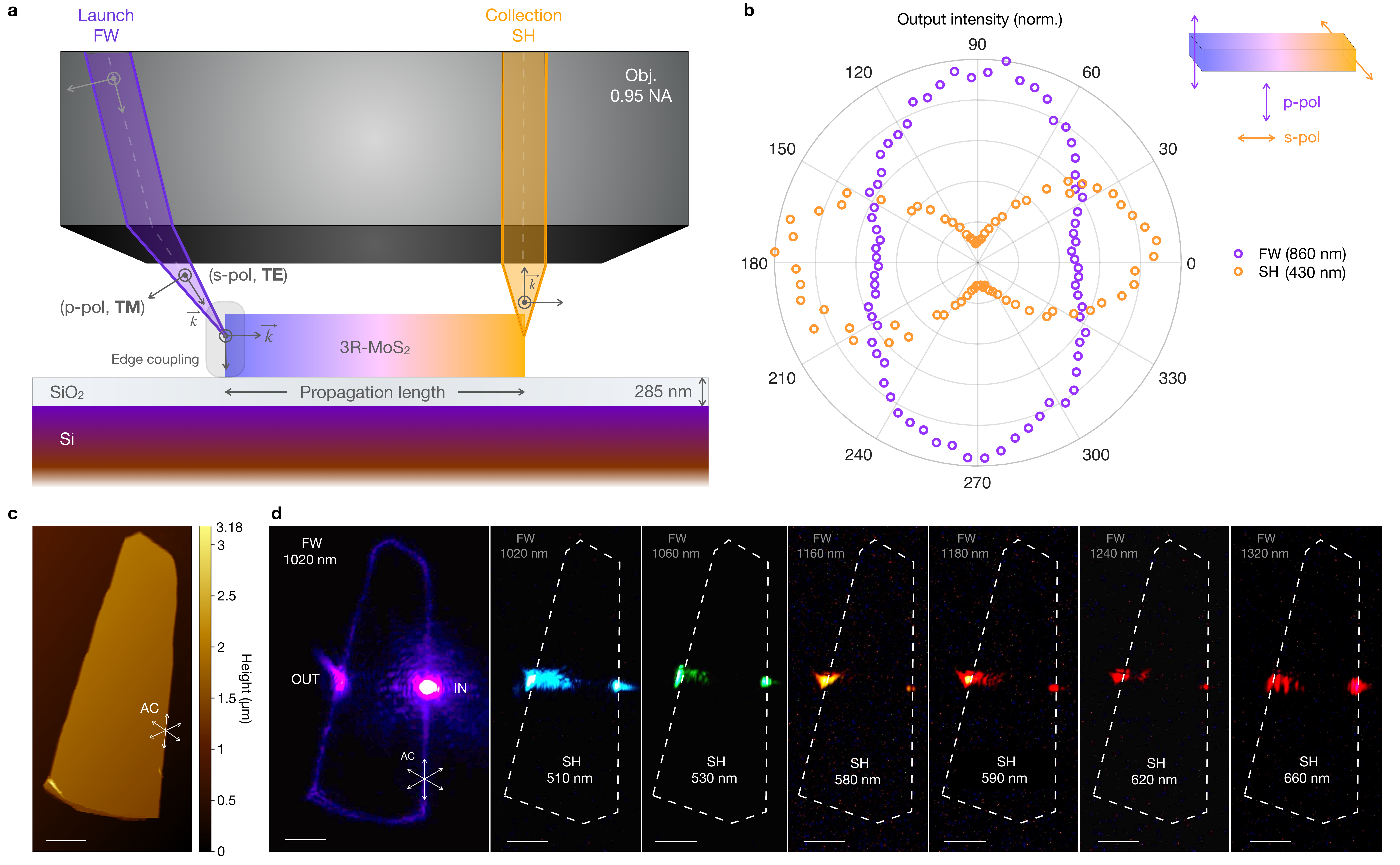}
\centering
\caption{\scriptsize{\textbf{Waveguide SHG in 3R-MoS$_2$.} \textbf{a,} Schematic diagram of the edge coupling in reflection geometry. The excitation beam (violet) is displaced away from the center of the objective to achieve edge coupling on one side of the 3R-MoS$_2$ flake, which has a thickness of $\sim \SI{1.2}{\micro m}$ and a lateral size of $\sim \SI{25}{\micro m}$ (propagation length). With the same objective we launch the FW and collect the emitted SH from the other side of the flake. \textbf{b,} Collected output intensity of FW and SH as a function of the input polarization. For p-polarized excitation we achieve the highest transmission of the FW, while the SH is maximum for s-polarized excitation. \textbf{c,} AFM map of the flake used as a waveguide. \textbf{d,} Images of the edge coupling at FW = 1020 nm and the broadly tunable SH fringes at different wavelengths. The dashed lines represent the edges of the sample. Scale bar: \SI{10}{\micro m}.}}
\label{fig:3}
\end{figure}

\noindent Increasing the nonlinear conversion efficiency of 3R-MoS$_2$ for propagation lengths beyond the coherence length requires phase matching, i.e. $\Delta k=0$. Phase-matched nonlinear interactions exploit the optical anisotropy (birefringence) of non-centrosymmetric nonlinear crystals. Notably, perfect phase matching achieved in waveguides lies at the heart of on-chip integrated nonlinear optics. In order to explore the birefringence of 3R crystals, in the following we show far-field edge coupling of the FW into a 3R-MoS$_2$ flake enables broadband SH emission in waveguide geometries, then we employ near-field imaging to visualize waveguided modes.

We use a confocal microscope in reflection geometry (Fig. \ref{fig:3}a) to probe the nonlinear frequency conversion in a waveguiding flake of 3R-MoS$_2$. The FW beam is displaced to the side of the objective (0.95 NA) in order to achieve edge coupling on one side of the flake. By tuning the polarization of the FW, we launch both transverse electric (TE) and transverse magnetic (TM)-like modes. The SH generated inside the 3R-MoS$_2$ waveguide over a propagation length of $\sim\SI{30}{\micro m}$ is detected from the opposite side of the flake with the same objective. The output FW and SH intensities both depend on the FW polarization (Fig. \ref{fig:3}b). While the most efficient FW edge coupling inside the waveguide is achieved for p-polarized light, i.e. TM modes, the conversion efficiency of SHG is maximum when the FW is s-polarized, i.e. when we launch TE modes. We ascribe this result to the asymmetry of the FW electric field in the TE mode. The field is aligned to MoS$_2$ sheets, and to the armchair direction specifically, whose dipole moment is also asymmetric. 

Figure \ref{fig:3}c reports the AFM map of the flake in which we achieve broadly tunable waveguided SHG. The micrographs of the edge coupling of a representative FW at 1020 nm and the SH at 510 nm, 530 nm, 580 nm, 590 nm, 620 nm and 660 nm are shown in Fig. \ref{fig:3}d. Here the FW polarization is set parallel to the AC direction, which is aligned to the input edge of the flake (AC directions are shown in the AFM map and the top left panel of FW = 1020 nm).

\begin{figure}[h!]
\includegraphics[width=1\textwidth]{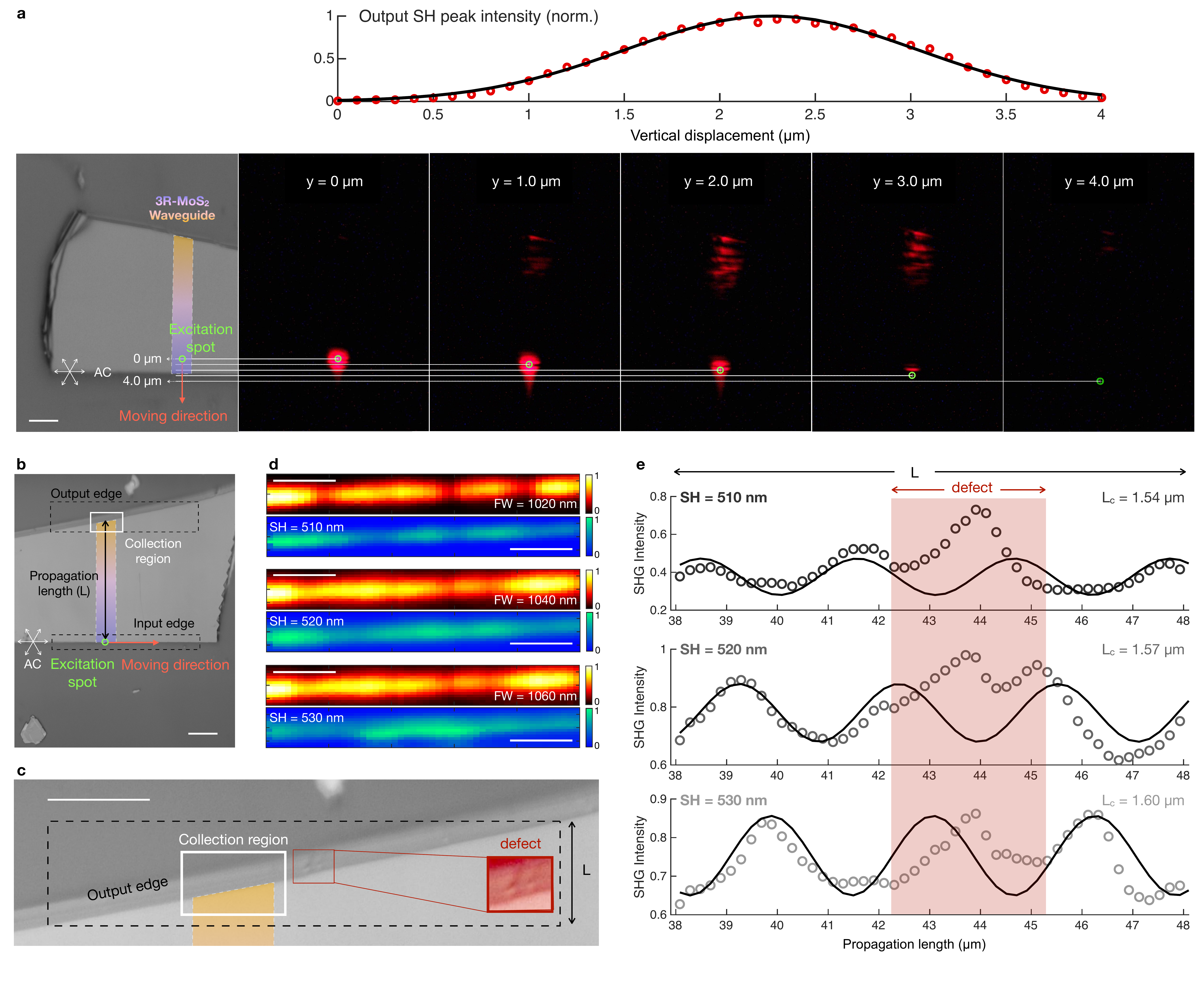}
\centering
\caption{\scriptsize{\textbf{Out-of-plane SHG coherence length in waveguide geometry.} \textbf{a,}  Image of the SH $(\SI{660}{nm})$ fringe pattern at different vertical coordinates of the FW  $(\SI{1320}{nm})$ excitation spot across the sample edge. The integrated SH intensity (top panel) at the output edge of the 3R-MoS$_2$ waveguide as a function of the vertical coordinate is fitted with a Gaussian profile. For all the reported 2D maps, the pump polarization direction is parallel to one of the AC directions, aligned to the sample edge. \textbf{b,} Optical image of the flake used for measuring SHG as a function of propagation length $L$, i.e. for the determination of the coherence length. \textbf{c,} Zoom-in on the output edge. All scale bars are $\SI{10}{\micro m}$. \textbf{d,} Transmitted FW/SH intensity maps at the output edge as a function of excitation spot coordinates across the bottom edge, at 3 different wavelengths. The scanned input area is $\SI{50}{\micro m} \times \SI{3}{\micro m}$. \textbf{e,} Normalized SH intensity as a function of $L$, along with the fitting curves and the extracted coherence lengths $L_\mathrm{c}$. The data under the shaded red region exhibit  deviations from the oscillating trend due to the present of a defect at output edge.}}
\label{fig:4}
\end{figure}

In Fig. \ref{fig:4}, we further investigate the mechanism of the edge coupling and the out-of-plane SH coherence length in 3R-MoS$_2$ waveguides. By vertically displacing the excitation spot across the input edge (Fig. \ref{fig:4}a), the SH fringe pattern changes accordingly, indicating that the FW coupling efficiency depends
sensitively on the relative position of the input edge. In this case, the overall intensity of the output SH fringe pattern as a function of the FW vertical displacement is fitted with a Gaussian profile, which is consistent with the approximate profile of the focused excitation.

To obtain the out-of-plane coherence length we measure waveguide SH as a function of the propagation length. We select a $\SI{775}{nm}$ thick 3R-MoS$_2$ flake (the AFM map is reported in Supplementary Figure 3) with a sharp horizontal input edge, and a diagonal output edge (Fig. \ref{fig:4}b-c). In this way, by scanning the FW beam along the input edge over a $\sim\SI{50}{\micro m}$ distance, we can collect the output FW and SH as a function of the propagation length within the slab. The intensity maps of FW and SH at different wavelengths are shown in Fig. \ref{fig:4}d. Upon scanning the FW beam along the input edge, at each point we collect the total transmitted FW and generated SH from the other side of the flake. The intensity of each pixel thus represents the total collected FW and SH, respectively, integrated over the collection region at the output edge. The measured FW maps quantify the actual FW intensity coupled into the flake, which can be affected by spatial inhomogeneities of the input edge. To quantify the thickness-dependent SHG with constant FW power, we normalize the SH intensity maps by the FW maps, as: SH / FW$^2$. The normalized SH intensity profiles at the 3 different wavelengths, as a function of the propagation length, i.e. the distance between input and output edges, are reported in Fig. \ref{fig:4}e. The SH intensity profiles are fitted to Eq. (2), with constant $I_{\omega}$. As expected, the region highlighted in red changes irregularly due to the presence of a defect at the output edge (see zoom-in of a spatial defect in the red box). The fitting profile of the oscillating phase-mismatched SHG provides the out-of-plane coherence lengths $L_\mathrm{c}$, which are \SI{1.54}{\micro m}, \SI{1.57}{\micro m} and \SI{1.60}{\micro m} at the SH wavelengths of 510 nm, 520 nm and 530 nm, respectively. Considering the multi-mode capacity of the 3R-MoS$_2$ in this thickness, the extracted $\Delta k$ here is likely related to the primary modes of the FW and SH with the mode dispersion relationship discussed in more details below. While in-depth optimization lies beyond the scope of this work, the waveguide frequency conversion and quantification of coherence lengths established here allow for future device fabrication, structuring and $\chi^{(2)}$ mode engineering in next-generation compact TMD platforms.

\begin{figure}[h!]
\includegraphics[width=1\textwidth]{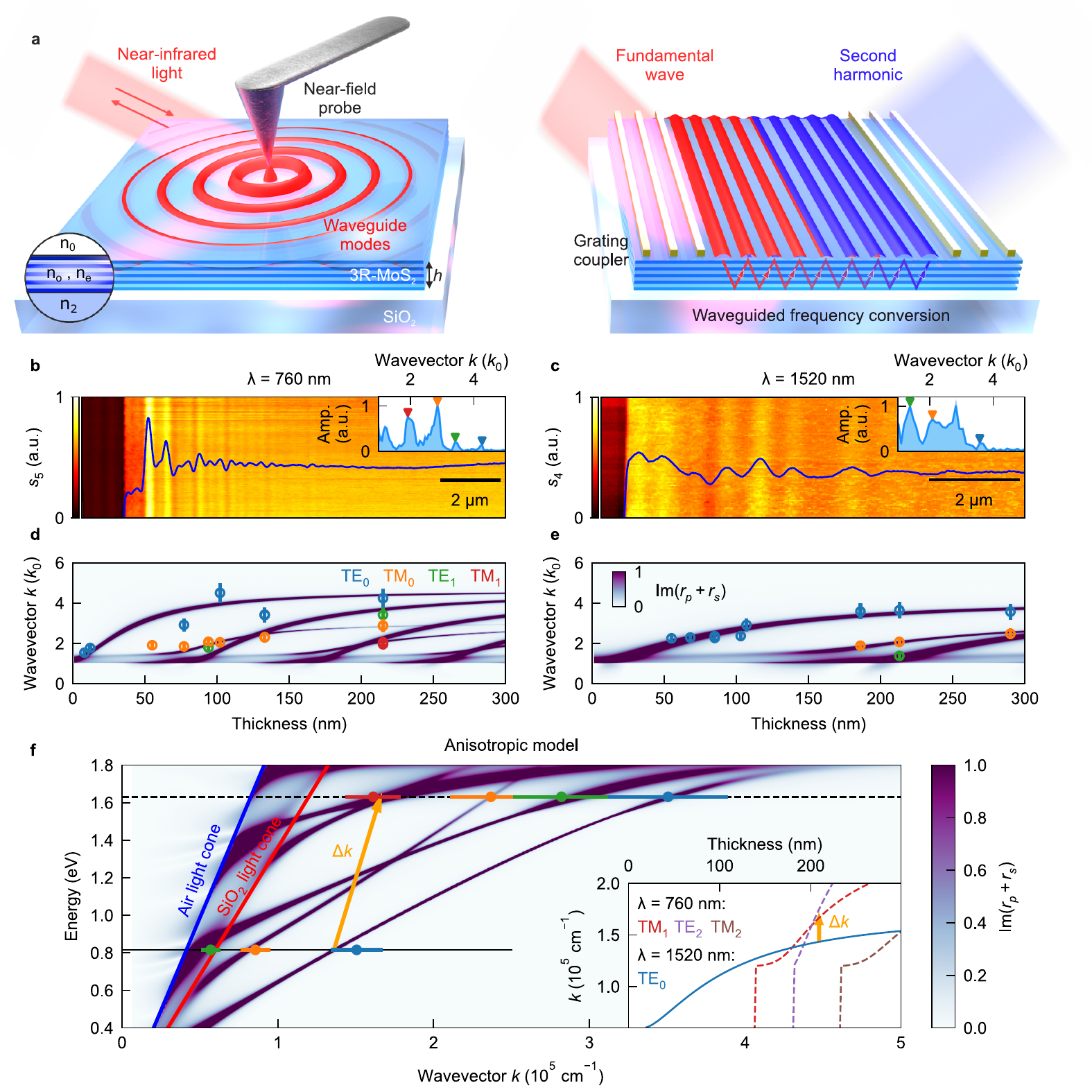}
\centering
\caption{\scriptsize{\textbf{Accessing the dispersion of waveguide modes (WMs) in 3R-MoS$_2$ via nano-imaging.} \textbf{a,} Schematics of the near-field experiments (left) and concept of grating coupled SHG in a TMD waveguide (right). \textbf{b,c,} Maps of the near-field amplitude $s_n$ obtained using excitation wavelengths $\lambda = \SI{760}{nm}$ (\textbf{b}) and $\lambda = \SI{1520}{nm}$ (\textbf{c}) on a flake with $h = \SI{215}{nm}$. The blue line was obtained by averaging along the vertical direction. Insets: Fourier analysis of the WMs (see Methods). The wavevector $k$ is given in units of the free-space wavevector $k_0$. \textbf{d,e,} Thickness dependence of the wavevectors $k$ of TE$_y$ and TM$_x$ modes at excitation wavelengths of $\lambda = \SI{760}{nm}$ (\textbf{d}) and $\lambda = \SI{1520}{nm}$ (\textbf{e}). The color code in \textbf{d} also applies to the symbols in \textbf{e} and in the insets of \textbf{b},\textbf{c}. The dispersion of the WMs is calculated via the imaginary part of Fresnel reflection coefficients for s-polarized  ($r_s$) and p-polarized ($r_p$) light. All error bars represent the relative uncertainty determined by the average FWHM of the peaks in the Fourier analysis. \textbf{f,} Anisotropic WM dispersion for $h=\SI{215}{nm}$ calculated with the same matrix formalism\cite{34} as in \textbf{d,e} and using the dielectric responses discussed in the main text as input. The yellow arrow indicates $\Delta k$ between different WMs for SHG at a FW of $\SI{1520}{nm}$ ($\SI{0.815}{eV}$). Inset: Phase-matching of WMs. The wavevector $k$ of the TE$_0$ mode (solid line) at the FW can be matched to higher-order SH WMs (dashed lines) at $\SI{760}{nm}$ by varying the crystal thickness and thereby minimizing $\Delta k$.}}
\label{fig:5}
\end{figure}

To further identify the conditions for phase matching, we characterize the birefringence of 3R-MoS$_2$ by imaging the propagation of waveguide modes (WMs) in real space using near-field nano-imaging. Due to their layered nature, van der Waals crystals exhibit vastly different dielectric properties along the in-plane and out-of-plane directions\cite{20,26}\noindent. Since the far-field experiments described above are mostly sensitive to the in-plane optical properties of thin 3R-MoS$_2$ flakes, in order to access the full dielectric tensor of 3R-MoS$_2$ we investigate the propagation of WMs\cite{10,26,27,28,29,30,31} featuring in- and out-of-plane electric field components using scattering-type scanning near-field optical microscopy\cite{32} (s-SNOM, see Fig. \ref{fig:5}a).

In maps of the scattered amplitudes $s_n$ (see Methods) at near-infrared photon energies, WMs manifest as periodic modulations. Figure \ref{fig:5}b shows the interference fringes close to the edge of a $\sim\SI{215}{nm}$ thick flake recorded with an incident wavelength of $\SI{760}{nm}$. The wavevectors of the contributing modes are shown in the inset of Fig. \ref{fig:5}b and were extracted with an established procedure (see Methods). Here, the momenta are given in units of the free-space wavevector $k_0$ of the incident light. In this case, the interference pattern comprises two transverse magnetic (TM$_x$) and two transverse electric (TE$_y$) modes, mostly characterized by out-of-plane and in-plane electric fields\cite{33}\noindent, respectively. Since the fields at the apex of the near-field tip are dominated by out-of-plane components, the TM$_x$ modes can be excited more efficiently and consequently have larger spectral amplitudes than the TE$_y$ counterparts. An analogous map of $s_n$ for an incident wavelength of $\SI{1520}{nm}$ is shown in Fig. \ref{fig:5}c.

To obtain the full refractive index tensor of 3R-MoS$_2$ for $\SI{760}{nm}$ and $\SI{1520}{nm}$, we systematically vary the sample thickness (see Fig. \ref{fig:5}d,e) and trace the evolution of TM$_x$ and TE$_y$ modes, thereby determining the in-plane and the out-of-plane refractive indexes, $n_o$ and $n_e$, respectively. We model the WMs dispersion via the imaginary part of Fresnel reflection coefficients for s-polarized ($r_s$) and p-polarized ($r_p$) light calculated with the code provided in ref.\cite{34}\noindent. We obtain the best agreement with our experimental data for: $(n_o,n_e)=(4.60,3.03)$ ($\lambda = \SI{760}{nm}$, see Fig. \ref{fig:5}d) and $(n_o,n_e)=(4.12,3.15)$ ($\lambda = \SI{1520}{nm}$, see Fig. \ref{fig:5}e). When the finite NA of the objective lens in the far-field experiment is considered, the near-field measurement of the in-plane dielectric response $n_o$ is consistent with the refractive index $(n)$ in Fig. \ref{fig:2}c. Due to the similar crystal structure, the in-plane properties of 3R-MoS$_2$ match previous reports on the 2H polytype\cite{26}\noindent. These results verify that infrared nano-imaging is a sensitive probe of anisotropic optical properties.

The full WM dispersion of a representative flake ($h\sim \SI{215}{nm}$) derived by the anisotropic model is provided in Fig. \ref{fig:5}f. Here, $n_o$ plotted in Fig. \ref{fig:2}c was used as an input and $n_e$ was kept constant – a reasonable assumption for the range of photon energies below the exciton resonances\cite{20} (compare Fig. \ref{fig:3}c). For this particular thickness $h$, the wavevector difference ($\Delta k$, previously visualized in the phase mismatch plot of Fig. \ref{fig:5}d) between WMs at the FW and at the SH is sizable. Due to the birefringence of the crystal, TM and TE branches exhibit significantly different dispersions. Therefore, by tailoring the thickness of the 3R-MoS$_2$ slab, the TE$_0$ modes at the FW and selected higher-order modes at the SH can be phase-matched in a waveguide geometry (see inset of Fig. \ref{fig:5}f). Different FWs or other nonlinear processes can be analyzed in a similar fashion. 

Finally, we note that edge coupling presented in Fig. \ref{fig:3} and Fig. \ref{fig:4} occurs at a natural edge of the flake. To achieve a more efficient in-plane momentum propagation through the waveguide, prism or grating couplers (see right panel of Fig. \ref{fig:5}a) directly placed on top of the waveguide would be beneficial. Further fabrication and structure engineering in this direction can allow for tailored mode excitations that will boost the conversion efficiencies of SHG in waveguides of van der Waals semiconductors.

\section*{Outlook and Conclusions}

\noindent We have fully characterized the second-order nonlinear frequency conversion from 3R-MoS$_2$, a naturally non-centrosymmetric layered material, as a function of the propagation length, both along the in-plane and the out-of-plane directions. In-plane SHG is generated by far-field normal incidence, while out-of-plane SHG is enabled by edge coupling in a waveguide geometry. We report both in-plane and out-of-plane SH coherence lengths, achieving a record value for the nonlinear conversion efficiency in TMDs, exceeding the monolayer value by more than four orders of magnitude. For nonlinear integrated photonics, our demonstration of waveguide SHG in 3R-MoS$_2$ slabs promises the same conversion efficiencies associated with LiNbO$_3$ but within propagation lengths that are two orders of magnitude shorter at telecom wavelengths\cite{21,35}\noindent. In addition,  waveguiding in van der Waals semiconductors will enable top-down fabrication compatibility and straight-forward integration to Si-based platforms. 

These results are fully corroborated by transfer-matrix calculations including both multilayer interference effects and phase-matching constraints. Furthermore, the full dielectric tensor of 3R-MoS$_2$ is accessed using waveguide-mode nano-imaging. The determined birefringence along in- and out-of-plane directions, as supported by numerical models, allows one to evaluate phase-matching conditions via mode dispersion relationship for any nonlinear process in a waveguide geometry as a function of sample thickness. Moreover, due to the larger transparency window along the out-of-plane direction of TMDs\cite{20}\noindent, it should be possible to harness the TM$_x$ modes, thereby partially circumventing the losses of the in-plane dielectric response close to the exciton resonances. This scheme provides a viable handle to design and evaluate integratable nonlinear photonic devices based on 3R TMD systems. 

In addition, due to the weak interlayer van der Waals forces, TMDs offer the key advantage of being easily stackable into vertical heterostructures with nearly arbitrary relative orientation or twist angle\cite{14} due to their atomically flat interfaces free of lattice mismatch limitations. This capability can be  exploited to extend the concept of quasi-phase-matching to non-centrosymmetric layered semiconductors using periodically poled TMD structures, achieved by stacking multilayer 3R-TMDs plates, each with a thickness corresponding to the coherence length determined in the present work - suitably rotated in order to introduce a $\pi$ phase shift between consecutive layers. Periodic poling in 3R-TMDs promises macroscopic nonlinear gain with values achieved in millimeter-thick crystals of standard materials, but with thicknesses that are more than 100-fold smaller. Thus, by virtue of the exceptional nonlinear properties and the possibility of cavity integration and phase-matching in waveguide geometries, we foresee ultra-compact devices with extremely high nonlinear conversion efficiency – even exceeding multi-pass state-of-the-art photonic resonators of aluminum nitride\cite{36} – opening new frontiers for engineering on-chip integrated nonlinear optical devices including periodically poled structures, photonic resonators, and optical quantum circuits.

\section*{ACKNOWLEDGEMENT}
\noindent The authors thank Aaron J. Sternbach for helpful discussions, Xingzhou Yan for experimental assistance. This work was supported by Programmable Quantum Materials, an Energy Frontier Research Center funded by the US Department of Energy, Office of Science, Basic Energy Sciences, under Award DE-SC0019443. C.T. and G.C acknowledge support by the European Union’s Horizon 2020 Research and Innovation program under Grant Agreement GrapheneCore3 881603. F.M. gratefully acknowledges support by the Alexander von Humboldt Foundation.

\section*{AUTHOR CONTRIBUTIONS}
\noindent X.X. and C.T. conceived the experiment and built the custom transmission microscope. X.X. prepared the samples and performed the nonlinear measurements. F.M. and S.Z. performed the near-field measurements. F.M. analyzed the near-field data and implemented the corresponding numerical models. Y.S. and X.X. determined the dielectric function from transmission/reflection experiments. G.C., D.B., and P.J.S supervised the study. X.X., C.T., F.M., G.C., D.B, and P.J.S. wrote the manuscript with input from all authors.

\section*{COMPETING INTERESTS}
\noindent The authors declare no competing interests.

\section*{DATA AVAILABILITY}
\noindent The data sets generated during and/or analyzed during the current study are available from the corresponding authors upon reasonable request.

\newpage
\section*{METHODS}

\textbf{Transmission spectroscope}
\newline
The custom-designed transmission microscope shown in Fig. \ref{fig:2}a is assembled with Cage System from Thorlabs Inc. The excitation laser is focused by a 40x reflective objective (Thorlabs) with numerical aperture NA=0.5. The emitted SHG and DFG are detected by a 50x objective (Nikon) with NA=0.95. The sample is loaded on a 3-axis piezo stage (PI)/2-axis manual stage (Thorlabs). The best focus of each flake is adjusted with the z-axis of the piezo stage while the position of the top/bottom objectives are fixed. The laser source (Coherent) is a Ti:Sapphire oscillator emitting $\SI{120}{fs}$ pulses at $\SI{1.55}{eV}$ with a repetition rate of $\SI{80}{MHz}$. The oscillator seeds an optical parametric oscillator emitting pulses tunable from $\SI{0.83}{eV}$ to $\SI{1.21}{eV}$. The excitation spot diameter on the sample is $\sim\SI{1}{\micro m}$, corresponding to a peak intensity of $\sim\SI{2.7}{GW/cm^2}$ for an average power of $\SI{1}{mW}$ impinging on the sample. The nonlinear emission is detected with a Silicon-EMCCD camera. Accounting for all the transmissive optical elements of the setup, both pump and signal pulses have a duration of $\sim\SI{250}{fs}$ at the sample plane, and in DFG mapping they are temporally synchronized by means of a mechanical delay stage before the excitation objective. \\

\textbf{Waveguide nano-imaging}
\newline
\noindent
Near-field experiments are performed with a scattering-type scanning near-field optical microscope (s-SNOM, Neaspec GmbH). The atomic force microscope (AFM) operates in tapping mode with a frequency of $\sim\SI{70}{kHz}$ and a tapping amplitude of $\sim\SI{50}{nm}$. The scattered light is detected using a photodiode and a pseudo-heterodyne scheme\cite{37}. To suppress any far-field background, the scattered amplitudes $s_n$ are additionally demodulated at higher harmonics of the tip tapping frequency.

\noindent
Based on this technique, WMs in multi-layer TMDs can be visualized as follows\cite{27,33}\noindent: continuous-wave radiation from a tunable Ti:sapphire laser\cite{38} is focused onto the metal tip (compare Fig. \ref{fig:4}a). There, the radiation is coupled into evanescent fields. As a result, this source of nano-light can excite WMs with momenta exceeding the light line, which subsequently propagate away from the tip apex as cylindrical waves. At the sample boundaries, the WMs are again coupled out into free space. Together with the incident light that is directly scattered from the tip, this radiation is collected by the parabolic mirror of the microscope. The interference of the light emerging from the tip apex and the sample edges gives rise to characteristic fringe patterns in maps of the scattered field amplitude $s_n$ (compare Fig. \ref{fig:4}b,c). Alternatively, the incident light can directly couple to WMs at the flake edges, propagate towards the tip, be scattered into the far field and interfere with radiation scattered directly from the tip. Nevertheless, both scenarios yield interference fringes with the same periodicity allowing for an extraction of the WM wavevector.\\

\textbf{Wavevector extraction and WM dispersion}
\newline
\noindent
In line traces of the scattered amplitude $s_n$ (compare lines in Figs. \ref{fig:4}b,c), the wavevectors of the WMs forming the interference pattern can be extracted via a Fourier transform. To this end, the spectral components generated by the step-like increase of $s_n$ at the sample edge need to be suppressed and the relative positions of tip, sample edge, and detector need to be taken into account. For the former, a Parzen window is used – a procedure introduced in ref.\cite{26}, whereas the geometrical correction derived in the Supplementary Information of Ref.\cite{27} is used for the latter. In short, the wavevector $k_{WG}$ of the WM is related to the observed wavevector $k_{Obs}$ given by the periodicity of the interference fringes via the following relation:

\[
    k_{WG} = k_{Obs}cos(\beta) + k_{0}cos(\gamma)sin(\beta+\delta)
\]
\noindent
Here, $\beta = sin^{-1}(\frac{k_0}{k_{WG}}cos(\gamma)cos(\delta))$, whereas $k_0$, $\gamma$, and $\delta$ are the wavevectors of the free-space radiation, as well as the out-of-plane and in-plane angles of incidence of the light with respect to the sample edge. For details, see ref.\cite{27}.
When considering the relative wavevectors $\frac{k_{WG}}{k_0}$ for the TM$_x$ and TE$_y$ modes, the dispersions in Figs. \ref{fig:4}d,e approach the out-of-plane ($n_e$) and in-plane refractive indices ($n_o$), respectively, in the limit of infinitely thick samples\cite{39}\noindent. As a result, the smaller values of $\frac{k_{WG}}{k_0}$  for $\lambda = \SI{1520}{nm}$ (Fig. \ref{fig:4}d) compared to the values for $\lambda = \SI{760}{nm}$ (Fig. \ref{fig:4}c) highlight a difference in refractive index even without further modelling.

\noindent
For a quantitative analysis of the WM dispersion, the matrix formalism provided in ref.\cite{34} was adapted to calculate the Fresnel reflection coefficients $r_s$ and $r_p$ for anisotropic multi-layered structures. For the data in the inset of Fig. \ref{fig:4}f, the transcendental equations in ref.\cite{26} were solved instead. This analogous procedure essentially yields curves that trace the maxima of $Im(r_p+r_s)$ as, for example, shown in Fig. \ref{fig:4}d-f, while neglecting the finite thickness of the SiO$_2$ ($\sim\SI{285}{nm}$) and hence the Si chip underneath.\\

\textbf{Broadband reflectance and transmittance measurements}
\newline
\noindent
The near-infrared and visible reflectance and transmittance spectra of 3R-MoS$_2$ flakes were measured using a Hyperion 2000 microscope coupled with a Bruker FTIR spectrometer (Vertex 80V). A tungsten halogen lamp was used as a light source covering a frequency range of 0.5 to $\sim\SI{2.5}{eV}$. Unpolarized light was focused on the sample using a x15 objective and the aperture size was set to be smaller than the sample dimensions. The reflectance and transmittance spectra are normalized to the bare substrate region. A Mercury-Cadmium-Telluride (MCT) detector and a Silicon detector were used for the near-infrared and visible range, respectively.


\newpage

\begin{center}
\large
{\bf References}
\end{center}
\begin{thebibliography}{10}
\newcommand{\enquote}[1]{''#1''}

%1
\bibitem{1} Ono, M. et al. Ultrafast and energy-efficient all-optical switching with graphene-loaded deep-subwavelength plasmonic waveguides. \textit{Nat. Photonics} \textbf{14}, 37–43 (2020).

%2
\bibitem{2} Li, C. Nonlinear optics: Principles and applications. (Springer Singapore, 2016).

%2-1
\bibitem{2-1} Klimmer, S., Ghaebi, O., Gan, Z. et al. All-optical polarization and amplitude modulation of second-harmonic generation in atomically thin semiconductors. \textit{Nat. Photon.} (2021). https://doi.org/10.1038/s41566-021-00859-y

%3
\bibitem{3} Sun, Z., Martinez, A. \& Wang, F. Optical modulators with 2D layered materials. \textit{Nat. Photonics} \textbf{10} 227–238 (2016).

%3-1
\bibitem{3-1} Yao, K. et al. Enhanced tunable second harmonic generation from twistable interfaces and vertical superlattices in boron nitride homostructures. \textit{Sci. Adv} \textbf{7}, eabe8691 (2021).

%3-2
\bibitem{3-2} Ehren M. et al. Ultrafast Electronic and Structural Response of Monolayer MoS$_2$ under Intense Photoexcitation Conditions. \textit{ACS Nano} \textbf{8}, 10734-10742 (2014).

%3-3
\bibitem{3-3} Ghazal H. et al. Single Nanoflake Hexagonal Boron Nitride Harmonic Generation with Ultralow Pump Power. \textit{ACS Photonics} \textbf{8}, 1922-1926(2021).


%4
\bibitem{4} Dinparasti Saleh, H. et al. Towards spontaneous parametric down conversion from monolayer MoS$_2$. \textit{Sci. Rep.} \textbf{8}, 3862(2018).

%5
\bibitem{5} Wang, Y., Jöns, K. D. \& Sun, Z. Integrated photon-pair sources with nonlinear optics. \textit{Applied Physics Reviews} \textbf{8},  011314(2021).

%6
\bibitem{6} Caspani, L. et al. Integrated sources of photon quantum states based on nonlinear optics. \textit{Light: Science and Applications} \textbf{6}, e17100 (2017).

%6
\bibitem{6-1} Lin, KQ., Bange, S. \& Lupton, J.M. Quantum interference in second-harmonic generation from monolayer WSe$_2$. \textit{Nat. Phys.} \textbf{15}, 242–246 (2019).


%7
\bibitem{7} Yao, K. et al. Continuous Wave Sum Frequency Generation and Imaging of Monolayer and Heterobilayer Two-Dimensional Semiconductors. \textit{ACS Nano} \textbf{14}, 708–714 (2020).

%8
\bibitem{8} Chen, H. et al. Enhanced second-harmonic generation from two-dimensional MoSe$_2$ on a silicon waveguide. \textit{Light: Science and Applications} \textbf{6}, e17060 (2017).

%9
\bibitem{9} Trovatello, C. et al. Optical parametric amplification by monolayer transition metal dichalcogenides. \textit{Nat. Photonics} \textbf{15}, 6–10 (2021).

%10
\bibitem{10} Wu, L. et al. Giant anisotropic nonlinear optical response in transition metal monopnictide Weyl semimetals. \textit{Nat. Physics} \textbf{13}, 350–355 (2017).

%11
\bibitem{11} Li, Y. et al. Probing symmetry properties of few-layer MoS$_2$ and h-BN by optical second-harmonic generation. \textit{Nano Letters} \textbf{13}, 3329–3333 (2013).


%11-1
\bibitem{11-1} Malard, L. M., Alencar, T. V., Barboza, A. P. M., Mak, K. F. \& De Paula, A. M. Observation of intense second harmonic generation from MoS$_2$ atomic crystals. \textit{Phys. Rev. B} \textbf{87}, 1–5 (2013).

%11-2
\bibitem{11-2} Wang G, Marie X, Gerber I, et al. Giant enhancement of the optical second-harmonic emission of WSe 2 monolayers by laser excitation at exciton resonances. \textit{Physical review letters} \textbf{114}, 097403 (2015).

%11-3
\bibitem{11-3} Nagler, P. et al. Giant magnetic splitting inducing near-unity valley polarization in van der Waals heterostructures. \textit{Nat. Commun.} \textbf{8}, 1551 (2017).

%12
\bibitem{12} Mennel, L. et al. Optical imaging of strain in two-dimensional crystals. \textit{Nat. Commun.} \textbf{9}, 516(2018).

%13
\bibitem{13} Robert W. Boyd \textit{Nonlinear optics, Fourth Edition} (Academic Press, 2020).

%13-1
\bibitem{13-1} Hsu, W.-T. et al. Second Harmonic Generation from Artificially Stacked Transition Metal Dichalcogenide Twisted Bilayers. \textit{ACS Nano} \textbf{8}, 2951–2958 (2014)

%14
\bibitem{14} Liu, F. et al. Disassembling 2D van der Waals crystals into macroscopic monolayers and reassembling into artificial lattices. \textit{Science} \textbf{376}, 903-906(2020).

%15
\bibitem{15} Wang, Q., Kalantar-Zadeh, K., Kis, A. et al. Electronics and optoelectronics of two-dimensional transition metal dichalcogenides. \textit{Nat. Nanotech} \textbf{7}, 699–712 (2012).

%16
\bibitem{16} Säynätjoki, A., Karvonen, L., Rostami, H. et al. Ultra-strong nonlinear optical processes and trigonal warping in MoS$_2$ layers. \textit{Nat Commun} \textbf{8}, 893 (2017).

%17
\bibitem{17} Shi, J. et al. 3R MoS$_2$ with Broken Inversion Symmetry: A Promising Ultrathin Nonlinear Optical Device. \textit{Adv.Mater.}  \textbf{29}, 1701486 (2017).

%18
\bibitem{18} Zhao, M. et al. Atomically phase-matched second-harmonic generation in a 2D crystal. \textit{Light: Science and Applications} \textbf{5}, e16131 (2016).

%19
\bibitem{19} Song, B. et al. Layer-Dependent Dielectric Function of Wafer-Scale 2D MoS$_2$. \textit{Adv.Optical Mater.} \textbf{7}, 1801250 (2019).

%20
\bibitem{20} Ermolaev, G. A. et al. Giant optical anisotropy in transition metal dichalcogenides for next-generation photonics. \textit{Nat. Commun.} \textbf{12}, 854 (2021).

%20-1
\bibitem{20-1} Chanyoung Yim et al. Investigation of the optical properties of MoS$_2$ thin films using spectroscopic ellipsometry. \textit{Appl. Phys. Lett.} \textbf{104}, 103114 (2014).

%21
\bibitem{21} Cai, L., Gorbach, A. v., Wang, Y., Hu, H. \& Ding, W. Highly efficient broadband second harmonic generation mediated by mode hybridization and nonlinearity patterning in compact fiber-integrated lithium niobate nano-waveguides. \textit{Sci. Rep.} \textbf{8}, 12478 (2018).

%22
\bibitem{22} Shoji, I., Kondo, T., Kitamoto, A., Shirane, M. \& Ito, R. Absolute scale of second-order nonlinear-optical coefficients. \textit{J. Opt. Soc. Am. B} \textbf{14}, 2268-2294 (1997).

%23
\bibitem{23} Liu, W. et al. Strong Exciton-Plasmon Coupling in MoS$_2$ Coupled with Plasmonic Lattice. \textit{Nano Letters} \textbf{16}, 1262–1269 (2016).

%24
\bibitem{24} Vaquero, D. et al. Excitons, trions and Rydberg states in monolayer MoS$_2$ revealed by low-temperature photocurrent spectroscopy.\textit{Communications Physics} \textbf{3}, 194 (2020).

%26
\bibitem{26} Hu, D. et al. Probing optical anisotropy of nanometer-thin van der waals microcrystals by near-field imaging. \textit{Nat. Commun.} \textbf{8}, 1471 (2017).


%27
\bibitem{27} Sternbach, A. J. et al. Femtosecond exciton dynamics in WSe$_2$ optical waveguides. \textit{Nat. Commun.} \textbf{11}, 3567 (2020).

%28
\bibitem{28} Kusch, P., Mueller, N. S., Hartmann, M. T. \& Reich, S. Strong light-matter coupling in MoS$_2$. \textit{Phys. Rev. B} \textbf{103}, 235409 (2021).

%29
\bibitem{29} Mrejen, M., Yadgarov, L., Levanon, A. \& Suchowski, H. Transient exciton-polariton dynamics in WSe$_2$ by ultrafast near-field imaging. \textit{Sci. Adv.} \textbf{5}, eaat9618 (2019).

%30
\bibitem{30} Li, Z. et al. High-Quality All-Inorganic Perovskite CsPbBr 3 Microsheet Crystals as Low-Loss Subwavelength Exciton–Polariton Waveguides. \textit{Nano Letters} \textbf{21}, 1822–1830 (2021).

%31
\bibitem{31} Hu, F. \& Fei, Z. Recent Progress on Exciton Polaritons in Layered Transition‐Metal Dichalcogenides. \textit{Adv.Optical Mater.} \textbf{8}, 1901003 (2020).

%32
\bibitem{32} Chen, X. et al. Modern Scattering‐Type Scanning Near‐Field Optical Microscopy for Advanced Material Research. \textit{Adv.Mater.} \textbf{31}, 1804774 (2019).

%33
\bibitem{33} Hu, F. et al. Imaging propagative exciton polaritons in atomically thin WSe$_2$ waveguides. \textit{Phys. Rev. B} \textbf{100}, 121301 (2019).

%34
\bibitem{34} Passler, N. C. \& Paarmann, A. Generalized 4 × 4 matrix formalism for light propagation in anisotropic stratified media: study of surface phonon polaritons in polar dielectric heterostructures. \textit{Journal of the Optical Society of America B} \textbf{34}, 2128 (2017).

%35
\bibitem{35} Thomas, J. et al. Quasi phase matching in femtosecond pulse volume structured x-cut lithium niobate. \textit{Laser and Photonics Reviews} \textbf{7}, L17-L20 (2013).

%36
\bibitem{36} Bruch, A. W. et al. \SI{17000}{\%W} second-harmonic conversion efficiency in single-crystalline aluminum nitride microresonators. \textit{Applied Physics Letters} \textbf{113}, (2018).

%37
\bibitem{37} Ocelic, N., Huber, A. \& Hillenbrand, R. Pseudoheterodyne detection for background-free near-field spectroscopy. \textit{Applied Physics Letters} \textbf{89}, 101124 (2006).

%38
\bibitem{38} Zhang, S. et al. Nano-spectroscopy of excitons in atomically thin transition metal dichalcogenides. \textit{Nat. Commun.} \textbf{13}, 542 (2022).

%39
\bibitem{39} Hu, D. et al. Tunable Modal Birefringence in a Low‐Loss Van Der Waals Waveguide. \textit{Adv.Mater.} \textbf{31}, 1807788 (2019).

%40
\bibitem{40} Malitson, I. H. Interspecimen Comparison of the Refractive Index of Fused Silica. \textit{J. Opt. Soc. Am., JOSA} \textbf{55}, 1205–1209 (1965).

%41
\bibitem{41} Harland Tompkins, Eugene A Irene \textit{Handbook of Ellipsometry} (William Andrew, 2005).

\end{thebibliography}
\end{document}